\documentclass[aps,prb,reprint,superscriptaddress,showpacs]{revtex4-1}
\usepackage{color}
\usepackage{graphicx}
\usepackage[american]{babel}
\usepackage[T1]{fontenc}
\usepackage{amsmath}
\usepackage{amssymb}
\usepackage{amstext}
\usepackage{amsthm}
\usepackage{latexsym}
\usepackage{verbatim}
\usepackage{natbib}
\usepackage{array}
\usepackage{color} 

\usepackage{ulem}   % For strikethrough

\def\214{Sr$_2$IrO$_4$}
\def\327{Sr$_3$Ir$_2$O$_7$}
\def\J{J$_{1/2}$}
\def\JJJ{J$_{3/2}$}
\def\Ef{$E_{F}$ }
\def\dxz{d$_{xz}$}
\def\dyz{d$_{yz}$}
\def\dxy{d$_{xy}$}
\def\kx{k$_{x}$}
\def\ky{k$_{y}$}

\usepackage[draft]{todonotes}   % notes showed

\usepackage{xspace}

\newcommand{\ket}[1]{\ensuremath{|#1\rangle}\xspace}

\begin{document}

\title {ARPES study of orbital characters, symmetry breakings and pseudogaps \\ in doped and pure Sr$_2$IrO$_4$ }
\vspace{1cm}

\author{Alex Louat}
\affiliation {Laboratoire de Physique des Solides, CNRS, Univ. Paris-Sud, Universit\'{e} Paris-Saclay, 91405 Orsay Cedex, France}

\author {Benjamin Lenz}
\affiliation {CPHT, CNRS, Ecole Polytechnique, IP Paris, F-91128 Palaiseau, France}
\author {Silke Biermann}
\affiliation {CPHT, CNRS, Ecole Polytechnique, IP Paris, F-91128 Palaiseau, France}
\affiliation {Coll{\`e}ge de France, 11 Place Marcelin Berthelot, 75005 Paris, France}

\author{Cyril Martins}
\affiliation {Laboratoire de Chimie et Physique Quantiques, Universit\'e Paul Sabatier Toulouse III, CNRS, 118 Route de Narbonne, 31062 Toulouse, France}

\author{Fran\c cois Bertran}
\author{Patrick Le F\`evre}
\author{Julien E. Rault}
\affiliation {Synchrotron SOLEIL, L'Orme des Merisiers, Saint-Aubin-BP 48, 91192 Gif sur Yvette, France}

\author{Fabrice Bert}
\author{V\'{e}ronique Brouet}
\affiliation {Laboratoire de Physique des Solides, CNRS, Univ. Paris-Sud, Universit\'{e} Paris-Saclay, 91405 Orsay Cedex, France}

\begin{abstract}
\214 is characterized by a large spin-orbit coupling, which gives rise to bands with strongly entangled spin and orbital characters, called \J~and \JJJ. We use light-polarization dependent ARPES to study directly the orbital character of these bands and fully map out their dispersion. We observe bands in very good agreement with our cluster dynamical mean-field theory calculations. We show that the \J~band, the closest to the Fermi level $E_{F}$, is dominated by d$_{xz}$ character along $k_x$ and d$_{yz}$ along $k_y$. This is actually in agreement with an isotropic \J~character on average, but this large orbital dependence in $\mathbf{k}$-space was mostly overlooked before. It gives rise to strong modulations of the ARPES intensity that we explain and carefully take into account to compare dispersions in equivalent directions of the Brillouin zone. Although the latter dispersions look different at first, suggesting possible symmetry breakings, they are found essentially similar, once corrected for these intensity variations. In particular, the pseudogap-like features close to the $X$ point appearing in the nearly metallic 15\% Rh-doped \214 strongly depend on experimental conditions. We reveal that there is nevertheless an energy scale of 30meV below which spectral weight is suppressed, independent of the experimental conditions, which gives a reliable basis to analyze this behavior. We suggest it is caused by disorder.

\end{abstract}

\date{\today}

\maketitle

\section{Introduction}

Most Mott insulating oxides display orbital degrees of freedom beyond a non-degenerate half-filled single-orbital model \cite{ImadaRMP98,Maekawa2004}.
These degrees of freedom complicate the analysis, as they open the possibility of various types of orbital orderings. 
Even in a seemingly classical example of a Mott insulator such as V$_2$O$_3$, the orbital occupations change at the metal-insulator transition\cite{ParkPRB00,PoteryaevPRB07}.
In layered perovskites, like ruthenates, there are three partially filled t$_{2g}$ orbitals, creating a hidden one-dimensional character despite the tetragonal structure. 
Namely, chains of \dxz~orbitals run along k$_x$~(and respectively \dyz~chains along k$_y$), which may trigger nesting instabilities\cite{SidisPRL99}. An exception are cuprates, where a half-filled d$_{x^2-y^2}$ band at the Fermi level describes the electronic structure quite well. 
Recently, \214 has become another possible example, because the spin-orbit coupling quite efficiently splits the three t$_{2g}$ bands into a half-filled non-degenerate $J_{eff}=1/2$ band and two filled $J_{eff}=3/2$ bands\cite{BJKimPRL08,MartinsPRL11}, called hereafter \J~ and \JJJ. The geometry of the \J~band is rather unusual, with strongly entangled spin and orbital characters.  
\begin{equation}
J_{1/2}^{\pm m_J} = \frac{\ket{d_{yz}, \mp \sigma} \pm i \ket{d_{xz}, \mp \sigma} \pm \ket{d_{xy}, \pm \sigma}}{\sqrt{3}}
\end{equation}  
where $\sigma$ is the spin and m$_J$ the pseudospin. 
This \J~character was confirmed experimentally \cite{KimScience09,BogdanovNatCom15}. 
\214 thus realizes new low-energy Hamiltonians which may exhibit new forms of magnetic ordering (see for instance Ref.~\onlinecite{JackelliPRL09}).

In this paper, we use angle-resolved photoemission spectroscopy (ARPES) to check directly the orbital character of the bands approaching the Fermi level in \214. This is possible thanks to selection rules associated to the light polarization, which modulate the intensity of the bands, depending on their symmetry with respect to mirror planes of the structure \cite{DamascelliRMP03}. We uncover a strong polarization dependence of the ARPES intensity, which seems counter-intuitive at first, as \J~has a very isotropic shape by construction. We explain through explicit density functional theory (DFT) calculations that a \J~band on the \214 lattice should indeed exhibit a well defined $\mathbf{k}$-dependence of the orbital character that is in very good agreement with our findings. This \lq\lq{}hidden\rq\rq{} orbital degree of freedom is often forgotten and places \214 at an intermediate situation, between cuprates and ruthenates. It could play a role in subtle symmetry breakings reported in iridates, either a time-reversal symmetry breaking in pure \214 \cite{JeongNatCom17,ZhaoNatPhys15} or density waves in doped compounds \cite{ChenNatCom18,ChuNatMat17}. We carefully take into account these intensity modulations to discuss possible intrinsic changes in the dispersion along equivalent directions of the Brillouin Zone (BZ), both for the pure and Rh-doped \214. We conclude that they are identical within at least 50 meV.

%== Fig. 1 : FS vs polarization =============================
\begin{figure*}[tbh]
\centering
\includegraphics[width=0.95\textwidth]{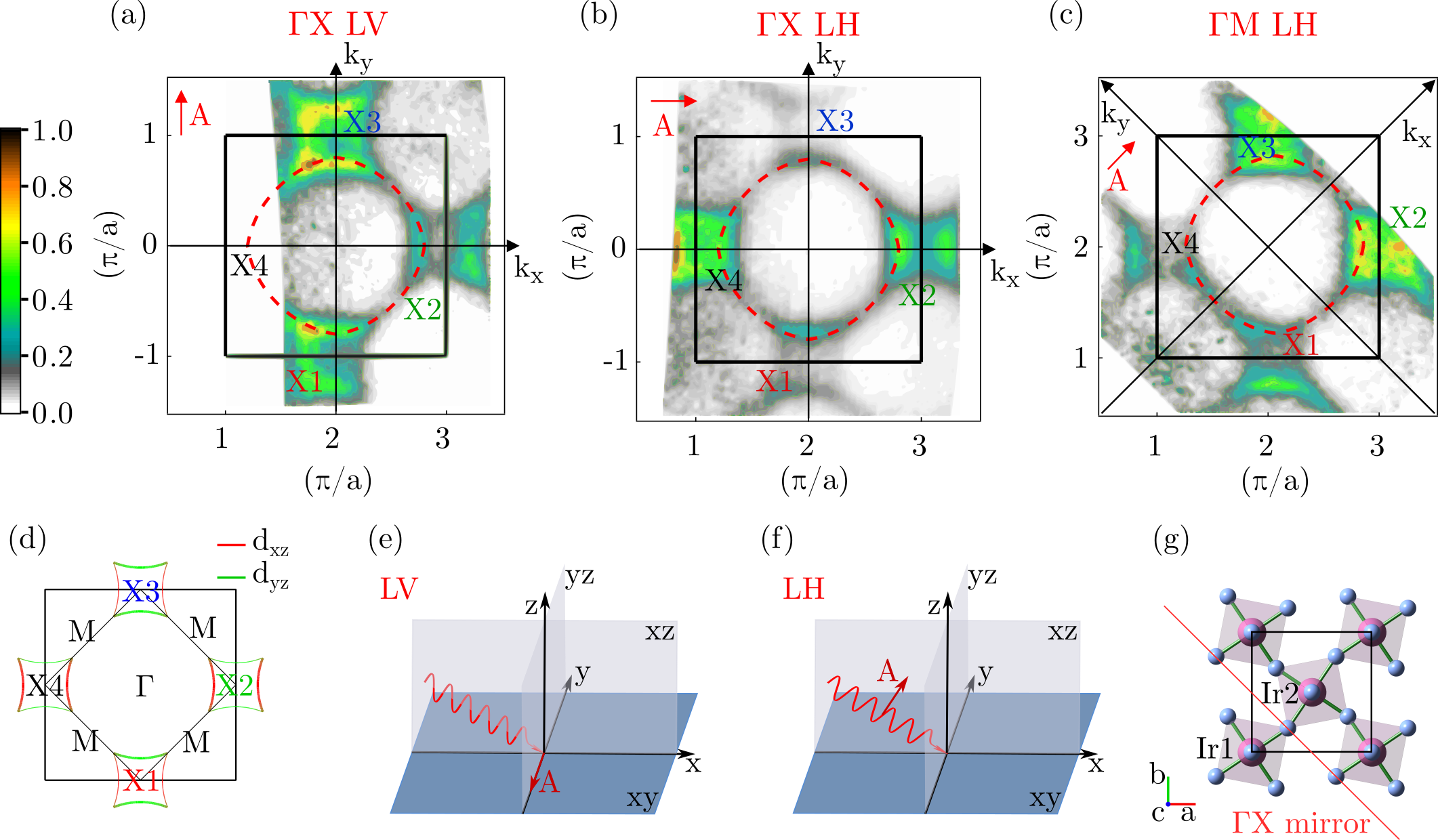}
\caption{ (a-c) Fermi Surface for 15\% Rh doped \214 at 50K in different experimental conditions : Linear Vertical (LV) polarization with $\Gamma$X oriented along $x$ (a), Linear Horizontal (LH) with $\Gamma$X oriented along $x$ (b) and LH with $\Gamma$M oriented along $x$ (c). The polarization A is indicated by the red arrow and the dotted red circle indicate the $\mathbf{k}$ positions of the spectrum closest to \Ef in this BZ. (d) Sketch of the Fermi Surface with hole pockets around X point having predominantly \dxz~or \dyz~character. Thin parts are folded with respect to the 2 Ir Brillouin zone [blue square, see also (g)].  (e-f) Geometry of our experiment for LV (e) and LH (f) configurations. (g) Sketch of the in-plane unit cell with the two Ir atoms and the $\Gamma$X mirror plane (blue circles represent oxygen atoms).}
\label{Fig_FS}
\end{figure*}
%\todo{CM:The dotted parts in Fig. 1 are difficult to see...}
%===============================

%==== Fig. 2 : Band structure along GMXG ===========================
\begin{figure*}[tbh]
\centering
\includegraphics[width=0.9\textwidth]{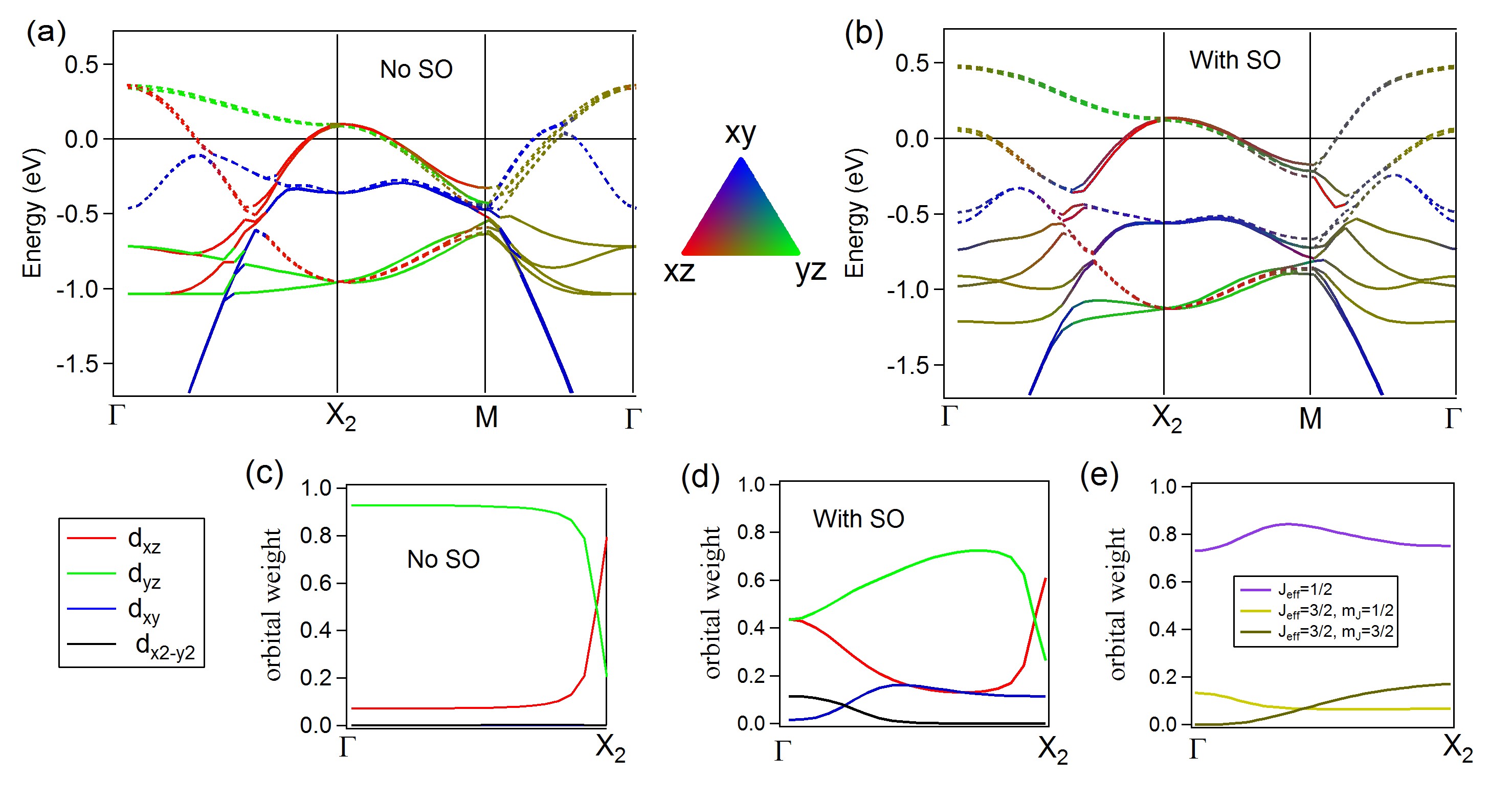}
\caption{Kohn-Sham band structures of \214\ within DFT-LDA without (a) and with (b) spin-orbit coupling. Direct and folded bands are plotted as solid and dotted lines, respectively. The $ t_{2g}$ orbital character obtained from the calculation is displayed using a three-color (RGB) scale : $d_{xz}$ in red, $d_{yz}$ in green and $d_{xy}$ in blue. 
 In (c) and (d), the decomposition of the orbital character for the top band along $\Gamma$X$_2$ is indicated, without and with spin-orbit, respectively. 
 In (e), the orbital weights for the same band in the J$_{1/2}$, J$_{3/2}$ basis are indicated.}
\label{Fig_GX}
\end{figure*}
%===============================
The possible role of hidden orders in the formation of pseudogaps has been much discussed, especially in the case of cuprates \cite{NormanPepinRepPhys03}. In iridates, two different kinds of pseudogaps have been reported. In electron doped systems, obtained by La \cite{DeLaTorrePRL15} or surface \cite{KimScience14} doping, the Fermi surface emerges from the M point (equivalent to the nodal point in cuprates), which may mean there is a pseudogap at the X point, similar to hole-doped cuprates. Theoretical calculations have discussed the possible emergence of a pseudogap due to strong antiferromagnetic fluctuations \cite{Martins2018,MoutenetPRB18}. In hole-doped iridates, which can be obtained by Ir/Rh substitutions \cite{ClancyPRB14}, the Fermi surface emerges around the X points \cite{CaoNatCom16,LouatPRB18}. A $\mathbf{k}$-dependent pseudogap was first reported \cite{CaoNatCom16}, implying some type of undefined symmetry breaking. In contrast, we have found a pseudogap over the entire Fermi surface (FS)\cite{LouatPRB18}, rather suggesting that it is a characteristic of the metallic state itself.%\todo{CM:I haven't seen (or understood) in part IV where we argue and conclude about this.}
In this paper, we clarify that the pseudogap-like features strongly depend on the experimental conditions, especially on polarization. 
This very likely explains the different conclusion of references~\onlinecite{CaoNatCom16} and~\onlinecite{LouatPRB18} .
We further show that at the X point spectral weight from unoccupied states leaks in and causes a leading edge in the electron distribution curve (EDC) at $\sim30$meV, which is independent of the experimental conditions. 
This finally gives a solid basis to analyze the pseudogap-like behavior in this particular case.

To compare dispersions and lineshapes between theory and experiment, we use an oriented cluster extension of dynamical mean-field theory (DMFT)\cite{Martins2018}.
%Going beyond single-site DMFT by including spatial fluctuations to treat the \J\ band is essential to obtain a correct spectrum in the paramagnetic phase. 
This technique has recently been applied successfully for pure and electron-doped \214 \cite{Martins2018,Lenz2019}. 
We present here results for the hole doped case, which simulates Rh doping. 
The Fermi surface and dispersions of all bands are in remarkable agreement with ARPES experiments. 
We do not find a pseudogap on the hole pockets, but intensity does extend to the X point. 
We then suggest that the pseudogap-like feature is associated with disorder due to in-plane Rh substitutions, which is not taken into account in the calculations.

\section{Orbital character in S\lowercase{r}$_2$I\lowercase{r}O$_4$}

\subsection{Polarization dependent ARPES measurements}

Fig. \ref{Fig_FS} (a-c) displays Fermi surfaces measured at 50K in \214 doped with 15\% Rh, at 100eV photon energy, with different polarization and orientation. The intensity modulations we are going to describe are however common to all dopings of \214. The intensity at X is due to the \J~band and is found stronger for X points along $k_y$ in (a), along $k_x$ in (b) and along the diagonal in (c), where the sample is rotated by 45$^\circ$. Obviously, different polarizations \lq\lq{}favor\rq\rq{} different X points, which implies a symmetry breaking that is not immediately expected from the equation given above for \J.

Well known selection rules\cite{DamascelliRMP03} state that orbitals odd (resp. even) with respect to a mirror plane will be detected with polarization odd (resp. even) with respect to that plane. The geometry of our experiment is sketched at the bottom of the figure. The light polarization A lies along the $y$ axis in linear vertical (LV) polarization and within the $xz$ plane in linear horizontal (LH) polarization (these sketches correspond to normal emission (\kx=0), the angle of A to the sample surface will change for higher \kx, see supplementary \cite{sup}). In LV, A is odd/$xz$ and even/$yz$, so that orbitals odd/$xz$ are expected to be detected along $k_x$ and orbitals even/$yz$ along $k_y$. The strong difference in intensities along $k_x$ and $k_y$ strongly suggests that we have bands even/$yz$ along $k_y$ (favorable case) and even/$xz$ along $k_x$ (unfavorable case). This is the case for \dyz\ (even/$yz$) and \dxz\ (even/$xz$), respectively. On the other hand, for LH, A is even/$xz$, while both components are present with respect to $yz$ \cite{sup}. While the interpretation of this situation is not as straightforward, the fact that the strongest intensity is now along \kx\ confirms that an orbital even/\kx\ dominates in this direction. 

We also observe a strong variation of intensity as a function of $k_x$ (more than a factor 10 between $k_x$=1 and $k_x$=3), indicating that the cross section increases when A is almost perpendicular to the surface (towards grazing incidence). This means that the orbitals at X must have a strong out of plane component, which is the case for \dxz\ or \dyz. 
  
If we now turn the sample by 45$^\circ$, the selection rules apply with respect to the plane containing the diagonal of the unit cell ($\Gamma$M) and the sample normal (this is not a true mirror plane, because of the oxygen positions). There is no difference anymore between the spots along $k_x$ and $k_y$, as expected for \dxz\ and \dyz\, which are equivalent with respect to this diagonal plane.% The main modulation of intensity comes for the angle between polarization and sample surface.

\subsection{Calculated orbital weights within DFT}

To understand the expected orbital characters observed around the X point, we calculate the $\mathbf{k}$-resolved spectral function and project it at each $\mathbf{k}$-point on the local Ir 5d-orbitals to get their orbital weights. In Fig. \ref{Fig_GX}, we show the Kohn-Sham band structure of \214\ calculated within the local density approximation (LDA)\footnote[1]{To distinguish \dxz~and \dyz~orbitals easier, we introduced a small orthorombicity in the structure ~(space group then becomes Fddd (70) with lattice constants $a$=7.775~\AA, $b$=7.776~\AA).} along $\Gamma$X$_2$M$\Gamma$ without spin-orbit coupling (SOC) in (a) and with SOC in (b). the calculation was performed using the WIEN2k software\cite{Wien2k}.We use as color scale the weight of each orbital \dxz~(red), \dyz~(green) and \dxy~(blue) obtained from this calculation and normalized by the total Ir weight. There is significant hybridization (nearly 50\%) with oxygen, but this does not affect the band symmetry. There is also significant hybridization with d$_{x^2-y^2}$ around $\Gamma$\cite{Martins2017}, that does not enter the color scale. 

There are two sets of $t_{2g}$ bands because of the 2 Ir per IrO$_2$ plane (see Fig. 1), with the two Ir orbitals in-phase or out-of-phase \cite{sup}. We call the first ones direct bands and write them (\dxz, \dyz, \dxy) [solid lines in Fig. \ref{Fig_GX}(a-b)], and the second ones folded bands (\dxz*, \dyz*, \dxy*) [dotted lines in Fig. \ref{Fig_GX}(a-b)]. The distinction between them is made simply by comparison to the bands of a non-distorted structure \cite{MartinsPRL11}. It is important to take this into account for ARPES measurements, as the two types of bands have opposite parities with respect to a $\Gamma$X mirror plane \cite{sup} and the folded bands have typically a much weaker intensity \cite{VoitScience00}. There is an additional doubling of the number of bands because of the two planes per unit cell, which gives a "bilayer splitting", which is usually small in \214 \cite{sup}, but it is clearly visible for the bands near -1eV at $\Gamma$ for example. 

%==== Fig. 3 : all dispersions + CDMFT, FS ===========================
\begin{figure*}[tbh]
\centering
\includegraphics[width=.9\linewidth]{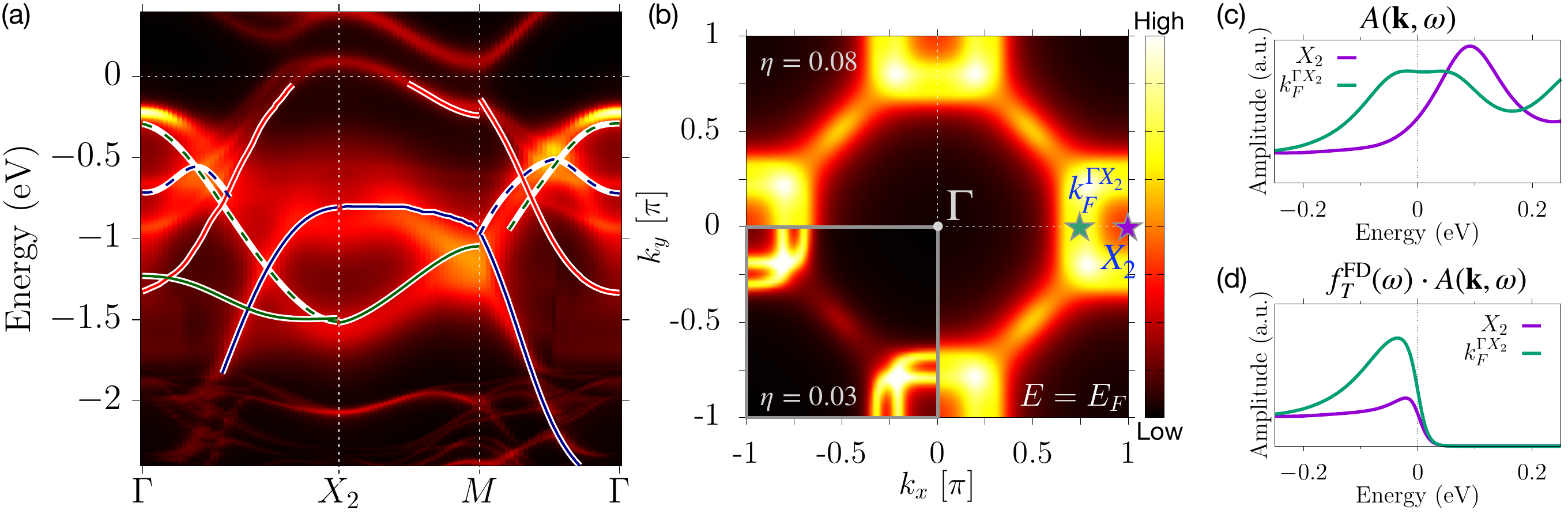}
\caption{(a) Extracted experimental dispersions (color lines) superimposed on an oriented cluster DMFT calculation with U=0.61eV and a hole density of $n_h=12\%$. Experimental dispersions were extracted in 15 \%Rh doped \214 and the raw data are shown in the supplementary. The colors are mainly there to follow the bands more easily and are just loosely connected to J=1/2 (one red band, the other band is above $E_F$), J=3/2 (green) and \dxy/d$_{x^2-y^2}$ (blue). (b) Corresponding Fermi surface for a broadening of $\eta=0.08$. The third quadrant shows the FS for $\eta=0.03$. (c) Spectral function $A(\mathbf{k},\omega)$ and energy distribution curves $f_{\mathrm{FD}}(\omega)\cdot A(\mathbf{k},\omega)$ around the Fermi energy at $X_2$ and $\mathbf{k}_F$ along $\Gamma-X_2$ as indicated by stars in (b).}
\label{Fig_OCDMFT}
\end{figure*}
%===============================

In Fig. \ref{Fig_GX}(c-e), we show explicitly the orbital character of the topmost band along $\Gamma$X$_2$. Without SOC, it is nearly of pure \dyz~character and meets a band of nearly pure \dxz~character near $\Gamma$ and X$_2$. At these points, the orbital character cannot be individually defined since the bands are degenerate. 
The SOC mixes the orbital characters [Fig. \ref{Fig_GX}(d)] and the colors of Fig. \ref{Fig_GX}(b) are indeed not as clear as in (a), but they are still dominated by one component in many places. One actually obtains a doublet of \J~character, as defined in Eq.(1), and a degenerate quartet of \JJJ~character, only when the SOC acts on three \textit{degenerate} $t_{2g}$ states. While the three t$_{2g}$ orbitals are nearly degenerate in \214 at the atomic level \cite{BogdanovNatCom15}, this is not the case for the Bloch states at a given $\mathbf{k}$ point. For instance, at the $\Gamma$ point, the bands of character \dxz* and \dyz* are degenerate at 0.3eV, but the \dxy* band is much lower around -0.5eV. Consequently, 
%the SOC splits the Bloch states at $\Gamma$ into an upper (folded) \J* doublet and a lower (folded) \JJJ* quartet whose components are separated by nearly 0.5eV. In addition, 
the character of the topmost band is mixed between \dxz\ and \dyz, with an admixture of only 10\% \dxy/d$_{x^2-y^2}$ instead of 33\% expected for a "standard" \J\ state. 
The situation is slightly different at X$_2$ since the \dyz* band lies near the Fermi level while the energy of \dxz* band is circa -1.0eV. As a result, the character expected near X$_2$ along \kx\ is now 70\% \dyz* for the topmost band. Note that the SOC acts only between bands of same nature (direct or folded) and is ineffective between direct and folded bands. Therefore, it does not lift the degeneracy between the bands of \dxz/ and \dyz* character. As the situation is reversed near X$_3$ along \ky\ (where the topmost band is of 70\% \dxz* character), the character of the whole band is in average over the Brillouin Zone essentially \J\ but at most of the $\mathbf{k}$-points, and notably at the X points, the character may be far from the standard (1/3,1/3,1/3) decomposition. This is confirmed in Fig 2(e) where we plot the orbital character of the topmost band along $\Gamma$X$_2$ with respect to the \J\ and \JJJ\ basis. The topmost band is on average composed of 80\% of \J~ character and of 20\% of \JJJ\ character along this direction, which is in good agreement with the decomposition given in Ref.~\onlinecite{JinPRB09}. %This $\mathbf{k}$-dependent deviation of the orbital character of the bands from the standard (1/3,1/3,1/3) \J\ model is often forgotten in discussions of \214, but is crucial to explain the observed modulations in intensity around the X points due to the polarization. This was recently used in another ARPES study \cite{ZwartsenbergCondMat19}.

The ARPES observation that a well-defined $t_{2g}$ orbital character persists at X (also used recently in ref. \onlinecite{ZwartsenbergCondMat19}) is then in agreement with the calculation. 
Introducing correlations beyond DFT will modify slightly the relative ratio of the decomposition at each $\mathbf{k}$-point. First, correlations enhance the spin-orbit polarization\cite{LiuPRL08,ZhouPRX18}, which brings the \JJJ\ band at $\Gamma$ below the Fermi level. Second, they open a gap at X in \214 to form an insulator. %This gap may partially close with Rh doping, but we suggested that metallicity mainly arises from hole doping with little effect on the gap \cite{LouatPRB18}.

\section{Comparison to DMFT calculations}

We now compare our data to a dynamical mean-field theory (DMFT) calculation, which captures both the gap opening and the correct position of \JJJ\ compared to \J. 
In Fig. \ref{Fig_OCDMFT}, we show a calculation for the hole doped case in which within the \J\ band the hole density is $n_h=12\%$. 
To treat the \J\ band within the calculation, we use an oriented cluster extension of DMFT, %, which crosses the Fermi level.
which has recently been applied successfully for pure and electron-doped \214 \cite{Martins2018,Lenz2019}.
Going beyond single-site DMFT by including spatial fluctuations to treat the \J\ band is essential to obtain a correct spectrum in the paramagnetic phase. 
The \J\ spectrum is then supplemented by the DMFT spectrum of the \JJJ\ manifold.
This assumes that the \JJJ\ spectrum does not change much upon hole doping due to its completely filled character, except for an additional energy shift due to the change in chemical potential.
We use an effective on-site Coulomb interaction of $U_{\mathrm{eff}}=0.6$eV as in the electron-doped case discussed in Ref.~\onlinecite{Martins2018}.

We have used different polarizations and geometries to measure all Ir bands of Rh-doped \214 from -2eV to the Fermi level E$_F$ \cite{sup}. The experimental dispersions (solid lines) are overlayed in Fig. \ref{Fig_OCDMFT}(a) onto the calculated spectral function. They compare well, in particular the \JJJ\ bands and the \J~ band features close to the Fermi level are well captured. Deviations seem to occur at the bottom of the \J~band, which is higher in the calculation. 
Comparing to the original DFT calculation shows that this is not due to a renormalization of \J\ induced by correlation. The problem is probably due to an inaccurate extraction of the experimental dispersion in this region, where two bands overlap \cite{sup}.

Comparing the calculated Fermi surface in Fig.~\ref{Fig_OCDMFT}(b) to the experimentally observed spectrum (Fig.~\ref{Fig_FS}) shows good agreement and motivates a detailed investigation of the spectrum.
In the following we focus on the spectrum at the high symmetry point $X_2$, as well as at the Fermi vector $\mathbf{k}_F$ along the path $\Gamma-X_2$.
Within the calculation, we have to choose a broadening $\eta$ of the spectral function. 
A value of $\eta=0.08$ leads to good agreement with experiment, but hides some of the FS features that become apparent at small broadening only. 
There, as in the electron-doped case\cite{Martins2018}, the Fermi surface is composed of two sheets due to the two different directions of antiferromagnetic fluctuations that are included via our cluster treatment.

%=== Fig. 4 : Dispersion pure/Rh ============================
\begin{figure*}[tbh]
\centering
\includegraphics[width=0.9\linewidth]{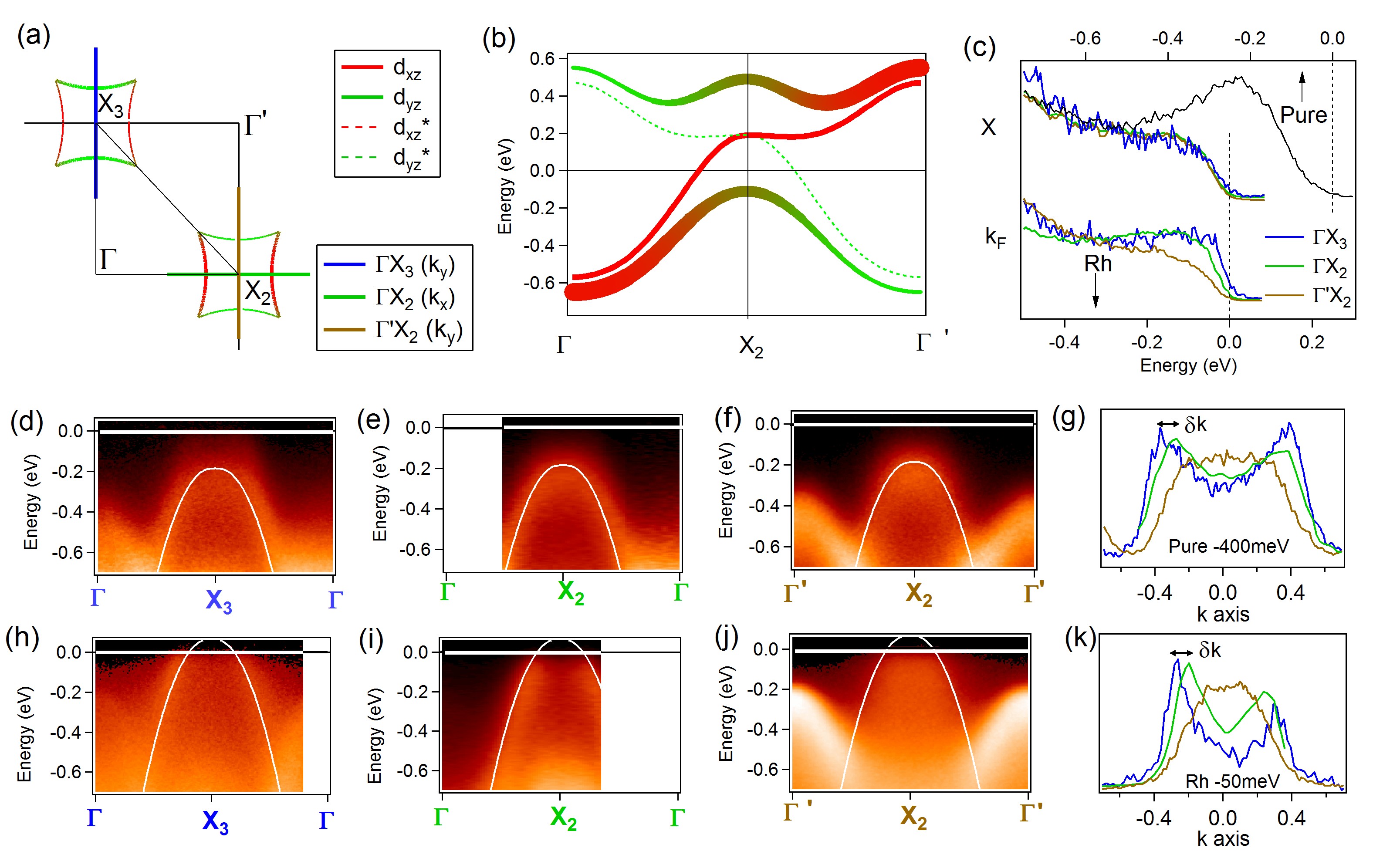}
\caption{(a) Sketch of the Fermi surface of Rh-doped \214 and positions of the different cuts represented below. (b) Sketch of the band structure along the path $\Gamma$-X$_2$-$\Gamma$\rq{}. Thin lines represent the \dxz~and \dyz* dispersion, before mixing. To simulate the gap opening, we assume a potential of 0.3eV along $\Gamma$X. The resulting dispersion is shown with markers having orbital character as color scale and weight as size (the direct band is assumed arbitrarily to have 10 times the weight of the folded band). (c) Energy distribution curve (EDC) spectrum for Rh doped sample at X (top) and $\mathbf{k}_F$ (bottom) for different $\Gamma$X directions. A reference spectrum at X for the pure case is shown in black with its abscissa (top) shifted by 0.25eV. This corresponds to the shift observed between the pure and Rh compound in the images (d-j). (d-f) Dispersion in \214 measured using LH polarization along $\Gamma$X$_3$ (d), $\Gamma$X$_2$ (e) and $\Gamma$\rq{}X$_2$ (f). The thin white line is a guide to the eye (the same in each case). (g) Momentum distribution curve (MDC) spectra at -0.4eV for the (d-f) cuts. (h-j) Same dispersions for 15\% Rh doped \214 measured in the same experimental conditions. The white dispersion is shifted up by 0.25eV. (k) MDC spectra at -0.05eV sample for the three cases (h-i).}
\label{Fig_disp}
\end{figure*}
%===============================

As can be seen in the top panel of Fig.~\ref{Fig_OCDMFT}(c), the spectral function at $X_2$ has a peak above $E_F$, as expected for a hole pocket around X. In the bottom panel of Fig.~\ref{Fig_OCDMFT}(c), we simulate the energy distribution curve (EDC) signal based on this data via $f^{\mathrm{FD}}_T(\omega)\cdot A(\mathbf{k},\omega)$, where $f^{\mathrm{FD}}_T(\omega)$ denotes the Fermi-Dirac function. The low energy tail of the spectrum gives the impression of a tiny peak at $E_F$ after multiplying with $f^{\mathrm{FD}}_T(\omega)$. The spectrum is quite different at the Fermi energy for $\mathbf{k}_F$. 
For the realistic broadening used in our calculation, the spectrum is relatively flat around $E_F$, see top panel of Fig.~\ref{Fig_OCDMFT}(c), but it has an underlying two-peak structure. After multiplying with the Fermi-Dirac distribution, the curve at $\mathbf{k}_F$ shows a peak close to $E_F$, but below. This should not be confused with a pseudogap, as the leading edge of the spectra (used experimentally to define the pseudogap\cite{sup}) is still defined by the Fermi edge.

\section{Impact of the intensity modulations on the ARPES lineshapes}

The polarization dependence of the ARPES intensity we have described is not only important to better understand the orbital structure of \J, but also because it strongly affects the intensity of the spectra and sometimes their shape. We will show that it is particularly important to take this into account to analyze pseudogap-like features around the $X$ point.

\subsection{Comparison of the dispersions along \kx~and \ky}

Figure~\ref{Fig_disp}(b) sketches the spectral function with its intensity modulations along the $\Gamma-X_2-\Gamma^{\prime}$ path. First, the dominant orbital along $\Gamma$$X_2$ will be \dxz, which have very different intensity in most experimental conditions compared to \dyz~along $\Gamma$$X_3$ (see intensity modulations in Fig. 1). Second, this dominantly \dxz~band joins at X$_2$ with a dominantly \dyz* band. These two bands are even with respect to the $\Gamma$X mirror plane\cite{sup}, so that they should have similar intensity under the same polarization, but the intensity of the \dyz* band is intrinsically much weaker due to its folded character\cite{VoitScience00}. 
When the gap opens at X$_2$, both the \dxz/\dyz* and direct/folded characters get mixed. The expected modulation of intensity along $\Gamma-X_2-\Gamma^{\prime}$ is illustrated by the size of the markers.

In the bottom part of Fig.~\ref{Fig_disp}, we compare the dispersion along three different $\Gamma$X paths for pure \214 [Fig.~\ref{Fig_disp}(d-f)] and Rh-doped \214 [Fig.~\ref{Fig_disp}(h-j)]. The distribution of intensity is very different along $\Gamma$X$_3$ (d) and $\Gamma\rq{}$X$_2$ (f), for example, which follows well the expectations of Fig. \ref{Fig_disp}(b). The intensity concentrates on the direct band, i.e. on the sides of the dispersion for (d) and at the center for (f). In Fig. \ref{Fig_disp}(e), the intensity is more equally distributed between the center and the sides, which is not expected from the direct/folded difference alone [(d) and (e) should be equivalent in that respect] and could be due to the contribution of \dxy~or imperfect selection rules. This nevertheless validates the existence of strong intensity variations near X.  

Besides these intensity variations, the dispersions themselves look different at first sight. They seem to extend closer to E$_F$ in (d) and (h). For the Rh case, it even looks as if there was no gap along $k_y$ (h) and a real gap along $k_x$ (i). Indeed, their leading edges at $\mathbf{k}_F$ [see Fig \ref{Fig_disp}(c)] are quite different, with an onset at larger energies for X$_2$ (35meV) than for X$_3$ (12meV). This could indicate a charge-density-wave-like gap opening along \kx~ and questions the equivalence of \kx~and \ky. To compare the dispersions more precisely, we overlay the same white curve over all images, just shifted up by 0.25eV for the Rh case, which is the estimated difference in chemical potential between the pure and the Rh-doped sample\cite{BrouetPRB15}. It is quite clear that the intensity is slightly outside this curve in (d) and (h) and inside in (e) and (f), suggesting again a truly different dispersion. This is illustrated more quantitatively by the momentum distribution curve (MDC) spectra taken at -0.4eV for the pure case [Fig. \ref{Fig_disp}(g)] and -0.05eV for the Rh case [Fig. \ref{Fig_disp}(k)]. In both cases, the peaks are found at slightly different positions along \kx\ and \ky\ with $\delta$k=0.06*$\Gamma$X. The dispersion through \ky~along $\Gamma$\rq{}X$_2$ looks however even more different, with peaks shifted towards X. This implies that a difference between \kx\ and \ky\ would not be a sufficient explanation. 

%=== Fig. 5 : Pseudogap ============================
\begin{figure*}[tbh]
\centering
\includegraphics[width=\linewidth]{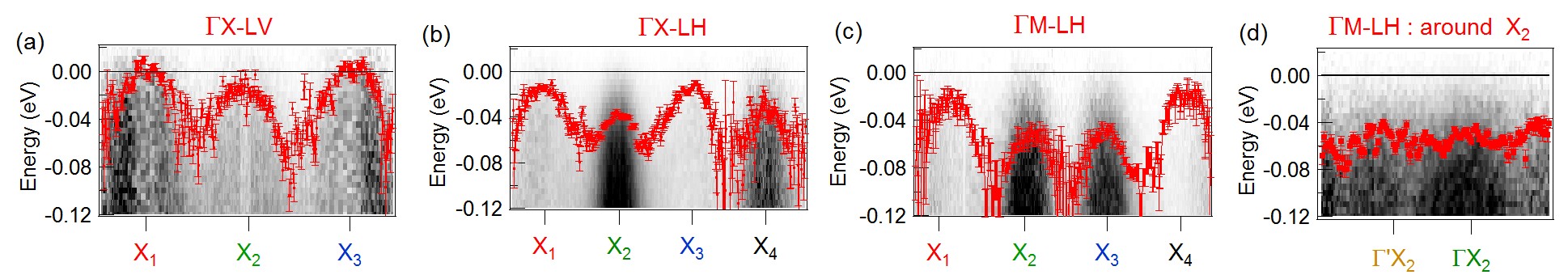}
\caption{(a-c) Leading edge spectra images taken along the circular contour, indicated as dotted red lines in Fig. 1(a-c). Red crosses indicate the position of the leading edge. (d) Leading edge spectra image around the X$_2$ point of the $\Gamma$M-LH FS in Fig. 1.} 
\label{Fig_PG}
\end{figure*}
%===============================

Two types of origins can be considered to explain such differences. It could be intrinsic due to a real splitting of the dispersion. As different experimental conditions may pick up different contributions of the two bands, this would lead to different peak positions. This splitting could be due to 1D fluctuations, as in the oriented cluster DMFT calculation described before, or other types of order, as the d-wave spin-orbit density wave \cite{ZhouPRX18}. The energy shift of the white curve needed to produce the observed splitting in momentum is of the order of 50meV. Alternatively, or additionnally, there could be an extrinsic contribution due to the intensity modulations. Such modulation may distort the lineshapes by giving different importance to contributions coming from different $\mathbf{k}$ points. As we know that there is indeed a strong intensity modulation around X, we give particular attention to this point. 

The most striking feature of the MDC in Fig. \ref{Fig_disp}(g) and (k) is actually that they are very asymmetric, distorted towards the center. Although a lorentzian is normally expected for a MDC curve\cite{DamascelliRMP03}, it is quite commonly distorted towards the occupied side of the band \cite{BrouetJElec12}. This can be understood as due to the integration of the tails of all EDC located at neighboring occupied $\mathbf{k}$ points. We show in supplementary that the change in intensity as a function of $\mathbf{k}$ described before can indeed satisfyingly explain the different shapes of MDC observed for the pure case. As the simulation depends both on the intrinsic shape of the spectral function (its width as a function of $\mathbf{k}$ for example, which is a priori unknown) and on the details of the extrinsic variation of intensity, we cannot simulate quantitatively the spectra. Therefore, we cannot exclude an intrinsic variation of the dispersion coexisting with this extrinsic modulation, with an upper bound of 50meV, however. This is not negligible, but the broad linewidths (0.23eV for the pure case shown in Fig. \ref{Fig_disp}(c) for example) do not allow to refine this further. On the other hand, this is already incompatible with the splitting of 0.2eV at X considered by Ref.~\onlinecite{ZhouPRX18} as a sign of d-wave spin-orbit density wave order. We note that such a splitting was also unclear in the raw data of Ref.~\onlinecite{DeLaTorrePRL15} and may be an artifact of the fitting procedure used in this paper. 

The case of Rh is more tricky as the band is now above the Fermi level at X and one would not expect a strong contribution to the lineshape from this unoccupied part. However, just like there is a remaining intensity at X coming from the tail of the spectra above E$_F$ in our calculation in Fig. \ref{Fig_OCDMFT}(c), there could be a sizable contribution in this case. Comparing with the spectra of the pure compound shifted up to the expected position for Rh (i.e. by 0.25eV, like the white dispersion) indeed suggests that some intensity can be expected at X. In this case, modulation of the relative intensity at $\mathbf{k}_F$ and at X could affect the lineshape and possibly change the leading edge shape. We now take a closer look at this point.

\subsection{Addressing the question of the pseudogap in R\lowercase{h}-doped S\lowercase{r}$_2$I\lowercase{r}O$_4$}

A pseudogap is usually defined as the position of the leading edge of the spectra closest to \Ef in one particular direction. In Fig. 1, we indicate by the dotted red circle the $\mathbf{k}$ points of such spectra and, in Fig.~\ref{Fig_PG}~(a-c), we build images with spectra taken along this contour. The full EDC spectra for the different $\Gamma$X directions are presented in supplementary. The leading edge is indicated by a cross. Values ranging from nearly zero to 60meV are found. It is clear that they are not associated to the different X points, but to the experimental conditions, as they change at a given X point for different experimental conditions. The tendency is to measure a larger pseudogap when the intensity is higher, wherever it occurs in $\mathbf{k}$-space. Further analysis is clearly needed to understand what should be called a pseudogap in this case.

The ARPES study of Ref. \onlinecite{CaoNatCom16} concluded that the pseudogap was different on the \lq\lq{}direct\rq\rq{} and \lq\lq{}folded\rq\rq{} parts of the Fermi Surfaces, forming \lq\lq{}Fermi arcs\rq\rq{} instead of hole pockets around X [see Fig. 1(d)]. This point is quite important to check, as it would completely change the FS topology. We note that the equivalence between the two sheets is dictated by periodicity of the 2 Ir unit cell itself, so that this proposal is quite puzzling. In Fig. \ref{Fig_PG}(d), we track the position of the leading edge along the hole pocket at X$_2$ in the $\Gamma$M-LH FS. We choose this case because the two sides have relatively similar intensities in this geometry. The leading edges of the spectra in Fig.\ref{Fig_PG}(d) are remarkably constant around -60meV. This rules out a specific pseudogap on the folded FS compared to the direct one. The different pseudogaps measured by Ref. \onlinecite{CaoNatCom16} are more likely due to different intensities on the two sheets in their measurement conditions, as for $\Gamma$X$_2$ and $\Gamma\rq{}$X$_2$ in Fig. \ref{Fig_disp}(c).

Interestingly, all spectra show the same leading edge at X [Fig. \ref{Fig_disp}(c)], despite the very different absolute intensities (more examples are given in supplementary \cite{sup}). However, their onset is located around -30meV from $E_F$, which is unexpected and anomalous for a metal. This defines a new energy scale to characterize this pseudogap-like behavior, independently of experimental conditions. As there are equal contributions of \dxz/\dyz~and direct/folded bands at X, it indeed makes sense that this measurement is less sensitive to the experimental conditions. On the contrary, the leading edge value measured at $\mathbf{k}_F$ is directly related to the relative intensity between X and $\mathbf{k}_F$. From our previous study of the pure case, we have seen that the relative intensity at X is strongest along $\Gamma$\rq{}X$_2$, then $\Gamma$X$_2$ and finally $\Gamma$X$_3$. This explains perfectly the trend of a larger leading edges onsets along $\Gamma$\rq{}X$_2$ (45meV), then $\Gamma$X$_2$ (35meV) and finally $\Gamma$X$_3$ (12meV). 

This shifts the discussion from the pseudogap-like feature to the meaning of this new energy scale at X. The intensity around X is incoherent, in the sense that there is no quasiparticle (QP) peak that could be defined anywhere with a width smaller than its binding energy. %We note that in iridates, quasiparticles have only been observed forming narrow electron pockets \cite{DeLaTorrePRL15,KimScience14,DeLaTorrePRL14,HeNatMat15} very close to $E_F$, typically within 30meV. Outside this region broad peaks, with dispersions unrenormalized compared to DFT calculations, dominate \cite{BrouetPRB18_327}. 
In the case of Rh, the disorder may enhance scattering between different $\mathbf{k}$ values. % and consequently the role of these incoherent features.
This would naturally explain why our calculations, which do not include any disorder effects, do not reproduce the experimentally observed leading edge position of the EDC at a non-zero binding energy.

\section{Conclusion}

Using polarization-dependent ARPES, we have directly determined the orbital character of the bands in \214. 
Despite the band dispersion being well described by the \J~picture, the band closest to the Fermi level has - at a given $\mathbf{k}$ point - well-defined orbital character in terms of cubic harmonics.
This is in line with \textit{ab initio} calculations and attaches great importance to anisotropies and orbital degrees of freedom.
From a conceptual point of view this observation has an interesting consequence:
the coefficients of an expansion of the \J~Wannier orbital in terms of cubic harmonics aquire $\mathbf{k}$-dependence.
This prevents the corresponding band from being spanned by a simple tight binding model with \textit{atomic} \J~orbitals. It is rather reminiscent of the Zhang-Rice construction in cuprate systems \cite{Zhang88}, where the Wannier orbital of dominant $d_{x^2-y^2}$ character is obtained as a superposition including $p_x$ and $p_y$ orbitals with $\mathbf{k}$-dependent coefficients.

In the ARPES experiment, the $\mathbf{k}$-dependent orbital variations give rise to strong matrix element effects that modulate the intensities.
Because of the $\mathbf{k}$-dependence of the \J~Wannier orbital itself the assumption of slowly varying matrix elements breaks down.
In addition, the presence of two transition metal atoms per unit cell, which is a rather common situation, imposes strong selection rules. As a consequence, dispersions measured in equivalent directions of the Brillouin zone sometimes appear different.
We considered both intrinsic (e.g. symmetry breakings) and extrinsic (e.g. matrix elements) origins for these differences. 
Although we cannot exclude the former, with an upper bound of 50meV to shifts or splittings that could be present in the electronic structure, we show that the strong modulation of ARPES intensity by matrix elements observed here is already able to explain most of these differences. 

In Rh-doped \214, we show that the leading edge of the spectra detected throughout the FS, commonly used to define a pseudogap, strongly depends on the experimental conditions. However, at the center of the hole pocket - a $\mathbf{k}$ point which is not part of the Fermi surface - spectral weight is observed with a leading edge of $30$meV, independently of experimental conditions. This establishes a new energy scale defining this highly anomalous metallic state. %We speculate its is due to disorder effects, as it is not present in our calculation. 

We have performed theoretical calculations within the recently developed oriented cluster dynamical mean-field scheme, which was already applied to electron doped \214.
Here, we show that it also provides a good description of the dispersion of hole-doped \214.
We provide a detailed comparison between the polarized ARPES measurements and the calculated spectral function. In theory, the
leading edge at the Fermi surface does not show a depletion of
spectral weight. We speculate that disorder effects, associated with Rh substitutions, are responsible for creating the spectral weight at X and the pseudogap-like feature. Rh-doped \214 appears as an interesting example of disordered correlated metal that deserves further studies.

More generally, our study indicates ways to better characterize pseudogap-like features with ARPES. Depressions of spectral weight at $E_F$ have indeed been observed in many correlated systems. Besides the famous example of cuprates, one dimensional systems \cite{GrioniJPCM09}, aperiodic crystals \cite{BrouetPRB13}, \lq\lq{}bad metals\rq\rq{} like manganites, nickelates and others \cite{NaamnehCondMat18} all tend to show a depression of intensity at the Fermi level. It is likely that all these behaviors do not have the same meaning and significance, but more precise definitions are lacking. A common feature of these situations is to exhibit broad lineshapes in energy and momentum. We emphasize in our study that this makes them sensitive to rapid intensity modulations and that non-lorentzian MDC lineshapes are a signature of such a situation. In this case, the simple estimation of the pseudogap by the leading edge value becomes insufficient and a full study of the intensity distribution is necessary.

%%%%%%%%%%%%%%%%%%%%%

\bibliography{Iridates_biblio}

\end{document}